\documentclass[12pt]{article}
\newcommand{\sect}[1]{\setcounter{equation}{0}\section{#1}}

\renewcommand{\appendix}{\setcounter{section}{0}
\renewcommand{\thesection}{\Alph{section}}}

\def\a{\alpha}

\def\e{\epsilon}

\def\P{\Psi}

\def\be{\begin{equation}}
\def\ee{\end{equation}}
\def\ba{\begin{eqnarray}}
\def\ea{\end{eqnarray}}

\newcommand{\nn}{\nonumber\\}
\newcommand{\no}{\nonumber}

\begin{document}
\renewcommand{\thefootnote}{\fnsymbol{footnote}}

\newpage
\setcounter{page}{0}
\pagestyle{empty}

\begin{center}
{\Large{\bf Multiplet with components of different masses\\}}
\vspace{1cm}
{\bf D.V. Soroka\footnote{E-mail: dsoroka@kipt.kharkov.ua} and
V.A. Soroka\footnote{E-mail: vsoroka@kipt.kharkov.ua}}
\vspace{1cm}\\
{\it Kharkov Institute of Physics and Technology\\
1 Akademicheskaya St., 61108 Kharkov, Ukraine}\\
\vspace{1.5cm}
\end{center}

\begin{abstract}
{\small\rm A principle possibility for the existence of a multiplet including 
the components with the different masses is indicated.
This paper is dedicated to the memory of Anna Yakovlevna Gelyukh (Kalaida).}



\end{abstract}

\newpage
\pagestyle{plain}
\renewcommand{\thefootnote}{\arabic{footnote}}
\setcounter{footnote}0

\sect{Introduction}

We start from the citation a very surprising (for us)
appraisal of supersymmetry~\cite{gl,va,vs,wz} given by Yury Abramovich Golfand
during the Conference "Supersymmetry-85" at Kharkov State University in
1985.  He said~\cite{g} that supersymmetry did not justify his hopes to find
a generalization of the Poincar\'e group such that every its representation
include the particles of different masses. Golfand and Likhtman
had missed their aim, but had instead found supersymmetry, every representation
of which contains the fields of different spins.

So, the problem was raised and requires its solution. In the present paper we
give a possible solution of the problem of the multiplet which components have
the different masses. We illustrate the solution on the example of the
centrally extended $(1+1)$-dimensional Poincar\'e algebra~\cite{gios1,gios2,cj,
ss,dss}.

\sect{Tensor extension of the Poincar\'e algebra}

In the paper~\cite{ss} the tensor extension of the Poincar\'e algebra in $D$ 
dimensions
\ba
[M_{ab},M_{cd}]=(g_{ad}M_{bc}+g_{bc}M_{ad})-(c\leftrightarrow d),\no
\ea
\ba
[M_{ab},P_c]=g_{bc}P_a-g_{ac}P_b,\no
\ea
\ba
[P_a,P_b]=Z_{ab},\no
\ea
\ba
[M_{ab},Z_{cd}]=(g_{ad}Z_{bc}+g_{bc}Z_{ad})-(c\leftrightarrow d),\no
\ea
\ba
[P_a,Z_{bc}]=0,\no
\ea
\ba\label{2.1}
[Z_{ab},Z_{cd}]=0
\ea
was introduced and its Casimir operators
\ba\label{2.2}
Z_{a_1a_2}Z^{a_2a_3}\cdots Z_{a_{2k-1}a_{2k}}Z^{a_{2k}a_1},\quad(k=1,2,\ldots);
\ea
\ba\label{2.3}
P^{a_1}Z_{a_1a_2}Z^{a_2a_3}&\cdots&
Z_{a_{2k-1}a_{2k}}Z^{a_{2k}a_{2k+1}}P_{a_{2k+1}}\cr\nn
+Z^{aa_1}Z_{a_1a_2}&\cdots&
Z_{a_{2k-1}a_{2k}}Z^{a_{2k}a_{2k+1}}M_{a_{2k+1}a},\quad(k=0,1,2,\ldots);
\ea
\ba\label{2.4}
\e^{a_1a_2\ldots a_{2k-1}a_{2k}}Z_{a_1a_2}\cdots Z_{a_{2k-1}a_{2k}},\quad2k=D
\ea
were constructed. Here $M_{ab}$ are generators of rotations, $P_a$ are 
generators of translations, $Z_{ab}$ is a tensor generator and 
$\e^{a_1\ldots a_{2k}}$, $\e^{01\ldots{2k-1}}=1$ is 
the totally antisymmetric Levi-Civita tensor in the even dimensions $D=2k$.

Generators of the left shifts with a group element $G$, acting on the function 
$f(y)$
\ba
[T(G)f](y)=f(G^{-1}y),\quad y=(x^a,z^{ab}),\no
\ea
have the form
\ba
P_a=-\left({\partial_{x^a}}+
{1\over2}x^b{\partial_{z^{ab}}}\right),\no
\ea
\ba
Z_{ab}=-\partial_{z^{ab}},\no
\ea
\ba\label{2.5}
M_{ab}=x_a{\partial_{x^b}}-x_b{\partial_{x^a}}
+{z_a}^c{\partial_{z^{bc}}}-{z_b}^c{\partial_{z^{ac}}}
+S_{ab},
\ea
where coordinates $x^a$ correspond to the translation generators $P_a$,
coordinates $z^{ab}$ correspond to the generators $Z_{ab}$ and
$S_{ab}$ is a spin operator. In the expressions (\ref{2.5})
$\partial_y\equiv{\partial\over\partial y}$.

\sect{Two-dimensional case}

In the case of the extended two-dimensional Poincar\'e algebra the Casimir 
operators (\ref{2.2}), (\ref{2.3}) and (\ref{2.4}) can be expressed as 
degrees of the following generating Casimir operators:
\ba\label{2.6}
Z=-{1\over2}\e^{ab}Z_{ab},
\ea
\ba\label{2.7}
C=P^aP_a+Z^{ab}M_{ba},
\ea
where $\e^{ab}=-\e^{ba}$, $\e^{01}=1$ is the completely antisymmetric
two-dimensional Levi-Civita tensor. The relations (\ref{2.5}) can be 
represented as
\ba\label{2.8}
P_t\equiv P_0=-\partial_t-{x\over2}\partial_y,
\ea
\ba\label{2.9}
P_x\equiv P_1=-\partial_x+{t\over2}\partial_y,
\ea
\ba\label{2.10}
J={1\over2}\e^{ab}M_{ab}=-t\partial_x-x\partial_t+S_{01},
\ea
\ba\label{2.11}
Z=\partial_y,
\ea
where $t\equiv x^0$ is a time, $x\equiv x^1$ is a space coordinate, 
$y\equiv z^{01}$ is a coordinate corresponding to the central element $Z$ and 
the space-time metric tensor has the following nonzero components
$g_{11}=-g_{00}=1$. 

The extended Poincar\'e algebra (\ref{2.1}) in this case can be rewritten
in the following form (see also~\cite{cj}):
\ba
[P_a,J]={\e_a}^bP_b,\no
\ea
\ba
[P_a,P_b]=\e_{ab}Z,\no
\ea
\ba\label{2.14}
[P_a,Z]=0,\quad[J,Z]=0
\ea
and for the Casimir operator (\ref{2.7}) we have the expression
\ba\label{2.15}
C=P^aP_a-2ZJ.
\ea

For simplicity let us consider the spin-less case $S_{01}=0$.
Then with the help of the relations (\ref{2.8})--(\ref{2.11}) 
we obtain a mass square operator
\ba\label{2.12}
M^2\equiv\partial_{xx}-\partial_{tt}={P_x}^2-{P_t}^2-JZ-{t^2-x^2\over4}Z^2,
\ea
where the notations $\partial_{xx}={\partial^2\over\partial x^2}$ and 
$\partial_{tt}={\partial^2\over\partial t^2}$ are used.

\sect{New coordinates}

By a transition from $t, x$ and $y$ to the new coordinates
\ba
x_{\pm}={t\pm x\over2},\no
\ea
\ba\label{3.1}
y_-=y-{t^2-x^2\over4},
\ea
we obtain the following expressions for the generators:
\ba\label{3.2}
P_-=\partial_{x_-}, 
\ea
\ba\label{3.3}
P_+=2x_-\partial_{y_-}-\partial_{x_+},
\ea
\ba\label{3.4}
J=x_-\partial_{x_-}-x_+\partial_{x_+},
\ea
\ba\label{3.5}
Z=\partial_{y_-},
\ea
where
\ba
P_{\pm}=P_x\pm P_t.\no
\ea
These generators satisfy the following commutation relations:
\ba
[P_+,P_-]=-2Z,\no
\ea
\ba
[J,P_{\pm}]=\pm P_{\pm},\no
\ea
\ba
[J,Z]=0,\no
\ea
\ba\label{3.6}
[P_{\pm},Z]=0.
\ea
We see that $P_{\pm}$ are step-type operators.

The Casimir operator (\ref{2.15}) in the new coordinates takes the form
\ba\label{3.7}
C=P_+P_-+Z-2ZJ
\ea
and the mass square operator is
\ba\label{3.8}
M^2=P_+P_-+Z-ZJ-x_+x_-Z^2.
\ea

\sect{Multiplet}

As a complete set of the commuting operators we choose the Casimir operators 
$Z$, $C$ and rotation operator $J$.  Let us assume that there exist such a 
state $\P_{z,j}(x_+,x_-,y_-)$ that
\ba\label{4.1}
P_-\P_{z,j}(x_+,x_-,y_-)=0,
\ea
\ba\label{4.2}
Z\P_{z,j}(x_+,x_-,y_-)=z\P_{z,j}(x_+,x_-,y_-),
\ea
\ba\label{4.3}
J\P_{z,j}(x_+,x_-,y_-)=j\P_{z,j}(x_+,x_-,y_-).
\ea
The equations (\ref{3.2}) and (\ref{4.1}) mean that $\P_{z,j}(x_+,x_-,y_-)$ 
independent on the coordinate $x_-$. Then, as a consequence of the relations 
(\ref{3.4}), (\ref{3.5}), (\ref{4.2}) and (\ref{4.3}), we come to the 
following expression for the state $\P_{z,j}(x_+,y_-)$:
\ba\label{4.4}
\P_{z,j}(x_+,y_-)=a{x_+}^{-j}e^{zy_-},
\ea
where $a$ is some constant.

For the states
\ba\label{4.5}
{P_+}^k\P_{z,j}(x_+,y_-),\quad(k=0,1,2,\ldots)
\ea
we obtain
\ba\label{4.6}
J{P_+}^k\P_{z,j}(x_+,y_-)={\cal J}{P_+}^k\P_{z,j}(x_+,y_-)
=(j+k){P_+}^k\P_{z,j}(x_+,y_-),
\ea
\ba\label{4.7}
M^2{P_+}^k\P_{z,j}(x_+,y_-)&=&{\cal M}^2{P_+}^k\P_{z,j}(x_+,y_-)\cr\nn
&=&[(k+1-j)z-x_+x_-z^2]{P_+}^k\P_{z,j}(x_+,y_-),
\ea
\ba\label{4.8}
C{P_+}^k\P_{z,j}(x_+,y_-)=z(1-2j){P_+}^k\P_{z,j}(x_+,y_-),
\ea
from which we have
\ba\label{4.9}
{\cal J}=j+k,
\ea
\ba\label{4.10}
{\cal M}^2=(k+1-j)z-x_+x_-z^2.
\ea
The states (\ref{4.5}) are the components of the multiplet.

By excluding $k$ from the relations (\ref{4.9}) and (\ref{4.10}), we
come to the Regge type trajectory
\ba\label{4.11}
{\cal J}=\a(0)+\a'{\cal M}^2
\ea
with parameters
\ba\label{4.12}
\a(0)=2j-1+x_+x_-z,
\ea
\ba\label{4.13}
\a'={1\over z}.
\ea

\sect{Conclusion}

Thus, on the example of the centrally extended $(1+1)$-dimensional Poincar\'e
algebra we solved the problem of the multiplet which contains the components 
with the different masses.

It would be interesting to construct the models based on such a multiplet.

Note that, as can be easily seen from the commutation relation
\ba
[\partial_{x^a}+A_a,\partial_{x^b}+A_b]=F_{ab},\no
\ea
where $A_a$ is an electromagnetic field and $F_{ab}$ is its strength tensor, 
the above mentioned extended $D=2$ Poincar\'e algebra (\ref{2.14}) is
arisen in fact when an ``electron'' in the two-dimensional space-time is moving
in the constant homogeneous electric field.

\section*{Acknowledgments}

One of the author (V.A.S.) would like to thank V.D. Gershun, B.K. Harrison,
 A. Mikhailov, D.P. Sorokin, I. Todorov and 
A.A. Zheltukhin for the useful discussions. The authors are especially grateful
to E.A. Ivanov for the interest to the work and for a set of valuable remarks.

\end{document}